\documentclass[%
reprint,
superscriptaddress,
amsmath,amssymb,
aps,
prb,
noeprint,
floatfix
]{revtex4-2}

\usepackage[pdftex]{graphicx}
\usepackage{bm}
\usepackage{amsmath}
\usepackage{textcomp}
\usepackage{xr-hyper}
\usepackage{hyperref}
\usepackage[all]{hypcap}
\usepackage{xcolor}
\usepackage{placeins}
\usepackage{titlesec}
\titlespacing{\subsubsection} {0pt}{1ex}{1ex}
\titleformat{\subsubsection}{\centering\fontsize{10pt}{20pt}\selectfont\itshape}{\thesubsubsection.}{0.5em}{}

\usepackage{svg}

\makeatletter
\newcommand*{\balancecolsandclearpage}{
  \close@column@grid
  \cleardoublepage
  \twocolumngrid
}
\makeatother

\begin{document}

\title{Pressure-tunable phase transitions in atomically thin Chern insulator MnBi$_2$Te$_4$}
\makeatletter
\let\newtitle\@title
\let\newauthor\@author
\let\newdate\@date
\makeatother

\author{Albin M\'arffy}
\affiliation{Department of Physics, Budapest University of Technology and Economics, M\H uegyetem rkp.\ 3., H-1111 Budapest, Hungary}
\affiliation{MTA-BME Correlated van der Waals Structures Momentum Research Group, M\H uegyetem rkp.\ 3., H-1111 Budapest, Hungary}

\author{Endre T\'ov\'ari}
\email{tovari.endre@ttk.bme.hu}
\affiliation{Department of Physics, Budapest University of Technology and Economics, M\H uegyetem rkp.\ 3., H-1111 Budapest, Hungary}
\affiliation{MTA-BME Correlated van der Waals Structures Momentum Research Group, M\H uegyetem rkp.\ 3., H-1111 Budapest, Hungary}

\author{Yu-Fei Liu}
\affiliation{Department of Chemistry and Chemical Biology, Harvard University, Cambridge, MA 02138, USA}
\affiliation{Department of Physics, Harvard University, Cambridge, MA 02138, USA}

\author{Anyuan Gao}
\affiliation{Department of Chemistry and Chemical Biology, Harvard University, Cambridge, MA 02138, USA}

\author{Tianye Huang}
\affiliation{Department of Chemistry and Chemical Biology, Harvard University, Cambridge, MA 02138, USA}

\author{L\'aszl\'o Oroszl\'any}
\affiliation{Department of Physics of Complex Systems, ELTE Eötvös Loránd University, H-1117 Budapest, Pázmány Péter sétány 1/A, Hungary}
\affiliation{Wigner Research Centre for Physics, H-1525 Budapest, Hungary}

\author{Kenji Watanabe}
\affiliation{Research Center for Functional Materials, National Institute for Materials Science, 1-1 Namiki, Tsukuba 305-0044, Japan}

\author{Takashi Taniguchi}
\affiliation{Research Center for Functional Materials, National Institute for Materials Science, 1-1 Namiki, Tsukuba 305-0044, Japan}

\author{Su-Yang Xu}
\affiliation{Department of Chemistry and Chemical Biology, Harvard University, Cambridge, MA 02138, USA}

\author{P\'eter Makk}
\email{makk.peter@ttk.bme.hu}
\affiliation{Department of Physics, Budapest University of Technology and Economics, M\H uegyetem rkp.\ 3., H-1111 Budapest, Hungary}
\affiliation{MTA-BME Correlated van der Waals Structures Momentum Research Group, M\H uegyetem rkp.\ 3., H-1111 Budapest, Hungary}

\author{Szabolcs Csonka}
\affiliation{Department of Physics, Budapest University of Technology and Economics, M\H uegyetem rkp.\ 3., H-1111 Budapest, Hungary}
\affiliation{MTA-BME Superconducting Nanoelectronics Momentum Research Group, M\H uegyetem rkp.\ 3., H-1111 Budapest, Hungary}

\date{\today}
\begin{abstract}
Topological insulators lacking time-reversal symmetry can exhibit the quantum anomalous Hall effect. Odd-layer thick MnBi$_2$Te$_4$ is a promising platform due to its intrinsic magnetic nature, however, quantization is rarely observed in it. Our magnetoresistance measurements in the antiferromagnetic phase indicate, instead of a quantum anomalous Hall insulator, the presence of a trivial insulator state likely due to disorder, while at high magnetic field a Chern insulator state appears. By applying hydrostatic pressure we are able to tune the magnetic interactions and the characteristic energy scales in the phase diagram. The trivial band gap is reduced, suggesting the role of disorder decreases with the compression of the layers.
\end{abstract}

\maketitle

\counterwithin*{equation}{part}
\stepcounter{part}
\renewcommand{\theequation}{\arabic{equation}}

\section{Introduction}
\label{sec:intro}
In the quantum anomalous Hall effect (QAHE) a material exhibits a quantized Hall resistance without an external magnetic field, which requires the presence of an intrinsic magnetization which breaks time-reversal symmetry \cite{Yu2010, Chang2023}. This effect arises in topological insulators with ferromagnetic order, leading to dissipationless chiral edge states while the bulk remains insulating. Spatially manipulating such spin-polarized modes are expected to be building blocks of novel quantum bits \cite{Ferreira2013}.

\begin{figure}[t!]
    \centering
    \includegraphics[width=\columnwidth]{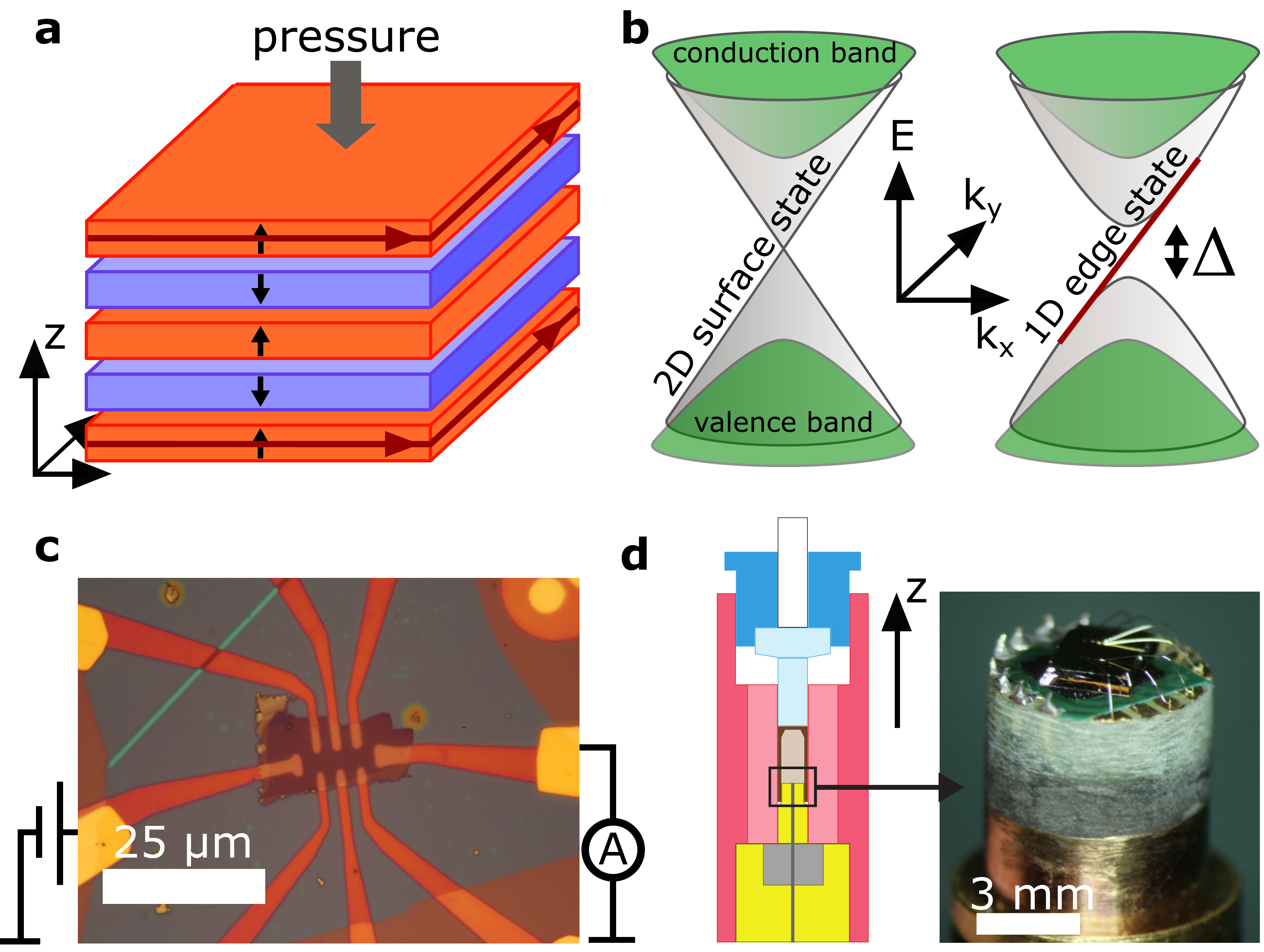}
    \caption{\label{fig:Dirac} (a) A-type AFM order for five SLs. Black arrows represent the SL magnetizations, and dark red arrows the chiral edge states. (b) Illustration of the surface state Dirac cone without (left) and with (right) a topologically non-trivial exchange gap. (c) Optical image of the device. Source and drain contacts are indicated, while the electrical setup of side contacts varied based on requirements. (d) A schematic of the pressure cell and a photo of the PCB carrying the chip.} 
\end{figure}

MnBi$_2$Te$_4$ (MBT) is a particularly promising material for the realization of the QAHE because it is an intrinsic magnetic topological insulator \cite{Zhang2019, Otrokov2019, Otrokov2019a}, as it naturally combines both topological and magnetic properties without requiring external doping. It is an A-type antiferromagnet (AFM) wherein the magnetization alternates across and is perpendicular to the septuple atomic layers (SL) \cite{Zhang2019, Ding2020}, as shown in Fig.\,\ref{fig:Dirac}(a). As van der Waals materials, MBT and related compounds are predicted to be easier to fabricate and control than topological materials doped with magnetic impurities, and they host a wide variety of topological quantum states \cite{Li2019,Liu2020}. Due to the exchange interaction related to the magnetization of the surface layers, the two-dimensional (2D) topological surface states become gapped as illustrated in Fig.\,\ref{fig:Dirac}(b). If the Fermi level is tuned into the gap, 1D chiral modes may propagate along the edges of the flake.

The characteristics of thin MBT films depend on the number of SLs and the magnetic phase \cite{He2020}. In the AFM phase, if the Fermi level is in the gap, odd (even)-SL MBT is expected to exhibit a quantum anomalous Hall (axion) insulator state with unity (zero) Chern number $C$, and the Hall resistance is quantized: $|R_\mathrm{xy}| = h/e^2$ \cite{Yu2010, Deng2020a, Chen2021a, Wang2024} (zero
\cite{Zhang2019, Otrokov2019a, Liu2020, Chen2021a, Li2023}) where $h$ is Planck's constant and $e$ is the elementary charge. In a high out-of-plane magnetic field $H$, the layers are fully polarized in a ferromagnetic (FM) phase, and $R_\mathrm{xy}$ is similarly quantized in a Chern insulator (CI) state with $|C|=1$, irrespective of the number of SLs \cite{Deng2020a, Liu2020, Ge2020, Liu2021, Ovchinnikov2021a, Gao2021, Cai2022, Ying2022, Li2024a}. 

However, the QAHE in odd-SL samples is often absent, and quantization may occur only in the field-polarized FM phase. Instead, a wide range of electronic properties has been observed at low field, for example a topologically trivial insulator state with near-zero $R_\mathrm{xy}$ \cite{Zhang2020, Ovchinnikov2021a, Ying2022, Cai2022, Chong2023a}. This discrepancy may be related to variations in material quality due to defects or surface degradation \cite{Zhao2021, Garnica2022, Tan2023, Li2024, Wang2024, Mei2024}, potentially affecting the magnetic interactions \cite{Yang2021, Chen2024}. Therefore, the goal of this study is to investigate the tunability of the magnetic interactions and the complex phase diagram by reducing the interlayer distance through applying pressure \cite{Fulop2021}. We performed magnetoresistance measurements on a 5-SL MBT sample which lacks the QAHE plateau. We studied phase transitions as a function of magnetic field and the band gaps through thermal activation measurements. We have found that the system is highly tunable with the pressure $p$. The trivial zero-field band gap is reduced by increasing $p$ while the CI band gap is weakly affected, moreover, the onset field of the FM phase increases. These observations are consistent with an increase in the interlayer AFM coupling, while the role of disorder appears to decrease with the compression of the layers.

\section{Experimental results}
\label{sec:exp}
The sample geometry is shown in Fig.\,\ref{fig:Dirac}(c). The device fabrication (detailed in Methods) was carried out in a glovebox. MBT was protected from air and solvents throughout the whole process. The chip was attached and wire-bonded to a PCB similarly to Ref.\,\citenum{Fulop2021}, then loaded into a hydrostatic pressure cell as demonstrated in Fig.\,\ref{fig:Dirac}(d). Measurements were carried out in a liquid helium cryostat equipped with a variable temperature insert, using low-frequency lock-in technique. The doped Si/SiO$_2$ substrate served as a gate electrode/dielectric. The pressure was applied at room temperature and set for each cool-down.

\begin{figure*}[!t]
\centering
    \includegraphics[width=\textwidth]{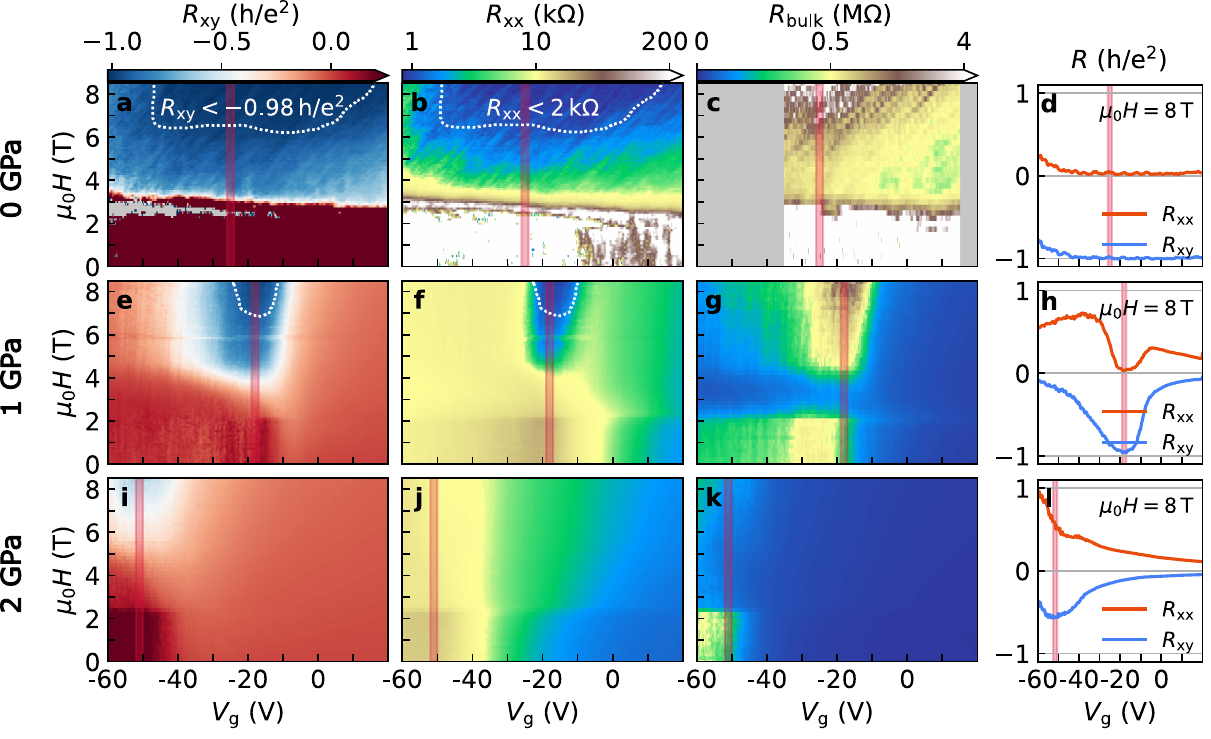}
    \caption{\label{fig:3x3} \textbf{Magnetoresistance at a series of pressures} (a) Map of the Hall ($R_\mathrm{xy}$), (b) the longitudinal ($R_\mathrm{xx}$) and (c) the bulk ($R_\mathrm{bulk}$) resistance as a function of gate voltage ($V_g$) and out-of-plane field $\mu_0H$ at 1.5\,K and 0\,GPa. (d) Corresponding horizontal cuts at $\mu_0 H = 8\,$T of $R_\mathrm{xx}$ (red) and $R_\mathrm{xy}$ (blue). (e-h) The same at 1\,GPa, and (i-l) at 2\,GPa. The CNP estimated at each pressure is marked by the red vertical lines. Areas outlined by white dashed lines in (a,e) mark where $R_\mathrm{xy}\leq -0.98\,h/e^2$ and in (b,f) where $R_\mathrm{xx} \leq 2\,\mathrm{k\Omega}$. For visibility, the color scale of $R_\mathrm{xx}$ and $R_\mathrm{bulk}$ has two linear slopes, with the center at $10\,$k$\Omega$ and at $0.5\,$M$\Omega$ respectively.} 
\end{figure*}

\subsubsection{Electronic phases}
First we discuss the gate voltage ($V_g$) and the external magnetic field ($H$) dependence of the Hall, longitudinal, and bulk resistances at pressures of approximately $p=0$, 1 and 2\,GPa, shown in Fig.\,\ref{fig:3x3}. The bulk resistance $R_\mathrm{bulk}$ was measured by grounding the contacts on both sides of the Hall bar (the three top and three bottom leads in Fig.\,\ref{fig:Dirac}(c)) between source and drain, effectively getting rid of potential edge state contributions to the current. 

Figs.\,\ref{fig:3x3}(b-c) plot $R_\mathrm{xx}$ and $R_\mathrm{bulk}$ as a function of $V_\mathrm{g}$ and $H$ at 0\,GPa. Their values up to about $\mu_0 H=3\,$T are in the range of hundreds of k$\Omega$ and several M$\Omega$, respectively. This demonstrates that the device is highly resistive in the studied gate range and that there are no edge states, as they would lead to an $R_\mathrm{xx}$ below a few k$\Omega$. The Hall resistance $R_\mathrm{xy}$, shown in Fig.\,\ref{fig:3x3}(a), reaches several $h/e^2$ in the same region (dark red area), which is caused by mixing with the large $R_\mathrm{xx}$ due to the irregular shape of the sample. These observations indicate a topologically trivial insulator state at low fields, similarly to Refs.\,\citenum{Zhang2020,Ovchinnikov2021a}. The insulating character is also supported by measurements of $R_\mathrm{xx}$ as a function of the temperature $T$ (see Fig.\,S4 in the Supplementary Material, SM).

In contrast, above around $\mu_0 H=6$\,T, the Hall signal becomes quantized, $R_\mathrm{xy} \approx -h/e^2$, while $R_\mathrm{xx}$ drops close to zero, consistent with a CI state. This region is marked by white dashed lines in Figs.\,\ref{fig:3x3}(a,b), and is also demonstrated in panel (d) as horizontal cuts at 8\,T. We estimate the position of the charge neutrality point (CNP) as approximately $-25$\,V and highlight it by vertical red lines in the figure. The presence of a CI state is supported by the increase in $R_\mathrm{bulk}$ at the CNP at high field (Fig.\,\ref{fig:3x3}(c)), as well as $T$-dependence (Fig.\,S4). In the regime between the trivial and Chern insulator states ($\sim 3-6$\,T) the reduced $R_\mathrm{bulk}$ suggests a closure or drop of the bulk band gap. The quantum Hall effect (QHE) can be ruled out as no fan-like features are visible on the colormap, and the sign of $R_\mathrm{xy}$ does not change with $V_\mathrm{g}$ at high field.

When the pressure is increased to $1\,$GPa, features similar to those described above are observed except in a narrower $V_\mathrm{g}$ range around a CNP of -18\,V, as demonstrated in Figs.\,\ref{fig:3x3}(e-h). At low field $R_\mathrm{xx}$ and $R_\mathrm{bulk}$ are relatively large, though not as high as at $0$\,GPa, while $R_\mathrm{xy}$ is on the order of $0.1\,h/e^2$. Including $T$-dependence in Fig.\,S5, these indicate that the system is still a trivial insulator. When increasing the field to above $\sim$2\,T, the bulk resistance exhibits a drop, then, after an intermediate regime, becomes high again. This is accompanied by a significant reduction in $R_\mathrm{xx}$, while $R_\mathrm{xy}$ reaches $-h/e^2$, heralding the appearance of the non-trivial (CI) state. These areas are again highlighted by white dashed lines. 
On either side of this $V_\mathrm{g}$ range, the Fermi level is tuned into the valence or conduction bands and the CI quantization disappears (Fig.\,\ref{fig:3x3}(h)). Compared to the $0\,$GPa case, the smaller width of insulating or quantized regions along the $V_\mathrm{g}$ axis suggests a decrease of the band gaps or a reduced density of defects. 

At $2\,$GPa (Figs.\,\ref{fig:3x3}(i-l)), the low-field trivial insulating state is still present with an even lower resistance (see also Fig.\,S6), while the CNP is shifted to -51\,V. A quantized $R_\mathrm{xy}$ and near-zero $R_\mathrm{xx}$ were not observed up to 8.5\,T. Nevertheless, there is a peak in $R_\mathrm{bulk}$ in the CNP at high field, and the features in $R_\mathrm{xy}$ are similar to the maps at lower pressures. Their tendencies with $V_\mathrm{g}$ and $H$ suggest that the FM phase and the corresponding CI state form at magnetic fields that are out of range of the measurements.

\subsubsection{Magnetic phases}

The longitudinal magnetoresistance and especially the anomalous Hall resistance make it possible to determine the magnetic phase transitions. We interpret them in the framework of a linear chain model\cite{Deng2020a, Ovchinnikov2021a, Yang2021, Bac2022, Chen2024} with the energy function
\begin{equation}
    \label{eq:linchain_main}
    E \propto H_E \sum_{j=2}^5 \bm{M}_j \bm{M}_{j-1} - \frac{H_a}{2} \sum_{j=1}^5 M_{j,z}^2
    - H \sum_{j=1}^5 M_{j,z},
\end{equation}
detailed in the SM. Here $H_E>0$ is the interlayer AFM exchange in units of A/m, $H_a>0$ is the anisotropy that defines the easy (out of plane, $z$) axis, and $\bm{M}_j$ are dimensionless layer magnetizations of unit length. 

\begin{figure}[!t]
    \centering
    \includegraphics[width=\columnwidth]{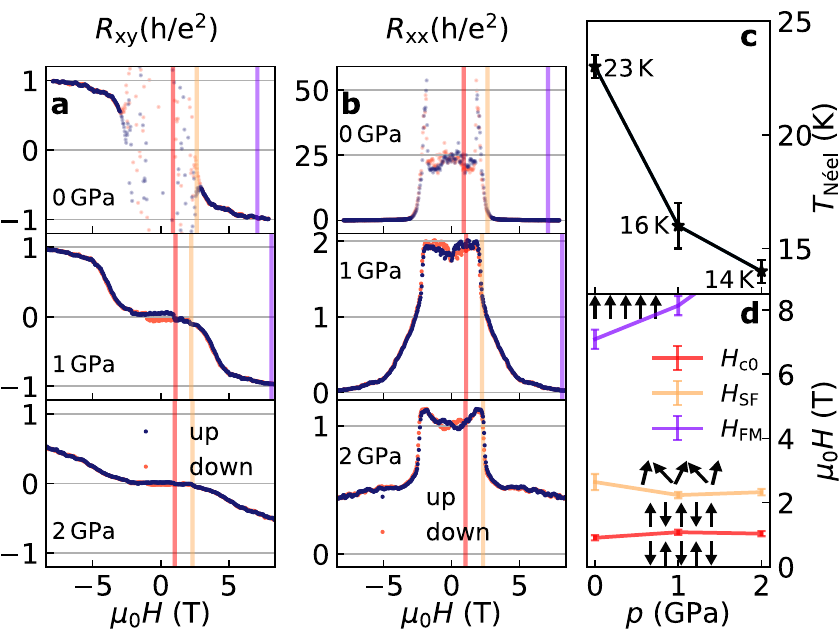}  
    \caption{\label{fig:bsymm} \textbf{Magnetic transitions.} (a) Up (black) and down (red) field-sweeps of the antisymmetrized Hall resistance and (b) the symmetrized longitudinal resistance (see Eqs.\,\ref{eq:antisymm1}, \ref{eq:antisymm2}) close to the CNP for all pressures at 1.5\,K. (c) Néel temperature vs pressure based on $R_\mathrm{xx}(T)$ or its derivative (see Fig.\,S2). (d) The estimated transition fields between the magnetic phases vs pressure, as highlighted by colored lines in (a,b). The orientation of SL magnetizations between them is illustrated by black arrows.}
\end{figure}

\begin{figure}[!t]
    \centering
    \includegraphics[width=\columnwidth]{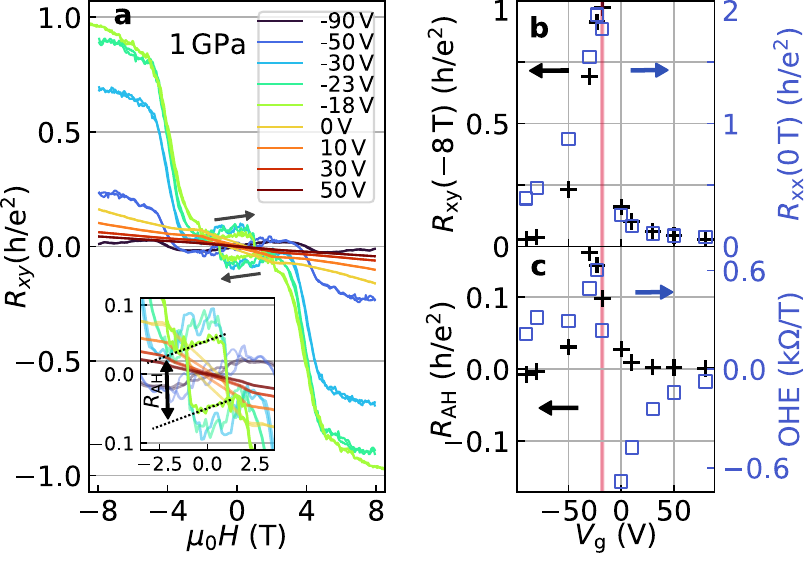}
    \caption{\label{fig:AHE} \textbf{AHE at 1 GPa and 1.5K.} (a) Antisymmetrized $R_{\mathrm{xy}}$ as a function of $H$ at a series of gate voltages. Inset: a zoom of the data, where the dotted lines are linear fits to one of the loops. (b) $R_{\mathrm{xy}}$ at -8\,T (black markers) and $R_{\mathrm{xx}}$ at 0\,T (blue) as a function of $V_g$. The CNP is marked by a red line. (c) The size $R_\mathrm{AH}$ of the hysteresis loop (black symbols) and the ordinary Hall coefficient (OHE, in blue) at low field.}
\end{figure}

We plot $H$-sweeps of the symmetrized longitudinal resistance $R_\mathrm{xx(S)}$ and the antisymmetrized Hall resistance $R_\mathrm{xy(AS)}$ close to the CNP in Figs.\,\ref{fig:bsymm} (b,a). Near zero field the system is in one of two, mirror-symmetric AFM states, illustrated in the bottom of Fig.\,\ref{fig:bsymm}(d). As the field is increased (black curves in Fig.\,\ref{fig:bsymm}), around $\mu_0 H = 1\,$T a small drop of $R_\mathrm{xy}$ can be observed at most pressures, highlighted by vertical red lines. This is part of a hysteresis loop between the up- and down-sweeps (black and red curves), and is most prominent at $1\,$GPa. The gate-dependence of its magnitude is plotted in Fig.\,\ref{fig:AHE}(a,c) and will be discussed further below. It is attributed to the first-order phase transition between the two AFM states, and is observable via the anomalous Hall effect (AHE). This complete flip of the magnetization of all SLs occurs at a field $H_{c0}$ when the Zeeman energy of the net magnetization becomes large enough compared to the AFM exchange $H_E$ and the anisotropy $H_a$. At $2\,$GPa a smaller hysteresis loop can be observed (see also Fig.\,S1(a,c) in the SM), while for 0\,GPa, its edge $H_\mathrm{c0}$ could only be determined through the conductivity $\sigma_\mathrm{xy}$ (displayed in Fig.\,S3) due to the contribution of a divergent and noisy $R_\mathrm{xx}$ to $R_\mathrm{xy}$. 

Starting from the energetically favorable AFM state above $H_\mathrm{c0}$, further increasing the field produces a suddenly sloping $R_\mathrm{xy}(H)$ and a drop in $R_\mathrm{xx}(H)$ for all pressures as highlighted by vertical orange lines ($H_\mathrm{SF}$) in Fig.\,\ref{fig:bsymm}(a,b). The transition is also easily identifiable in the non-symmetrized maps of $R_\mathrm{bulk}$ in Fig.\,\ref{fig:3x3} near $2-3\,$T. It is attributed to the spin-flop transition to the canted antiferromagnetic (cAFM) state, which is illustrated by the tilted arrows in Fig.\,\ref{fig:bsymm}(d). Here all layer magnetizations are partially aligned with the magnetic field as it dominates the anisotropy, but it still competes with the exchange $H_E$.

Above 7\,T, $R_\mathrm{xy}$ and $R_\mathrm{xx}$ saturate around $-h/e^2$ and zero, respectively, as shown in Fig.\,\ref{fig:bsymm}(a,b) for 0 and 1\,GPa. We highlight these fields ($H_\mathrm{FM}$) by vertical purple lines and attribute them to the onset of FM order and the CI state \cite{Deng2020a, Ovchinnikov2021a}. The transition fields between the phases are plotted in Fig.\,\ref{fig:bsymm}(d) with colors corresponding to the vertical lines in (a,b). The onset field of the FM phase ($H_\mathrm{FM}$, purple symbols) is much larger than that of the cAFM phase ($H_\mathrm{SF}$, orange), and goes out of range at 2\,GPa. 

We extracted the Néel temperature $T_\mathrm{N}$, which is revealed by a local maximum in $R_\mathrm{xx}(T)$ or can be determined from its derivative (see the SM and Fig.\,S2), and we show the results in Fig.\,\ref{fig:bsymm}(c). The estimated 0-GPa value of $23$\,K closely matches the values in the literature \cite{Otrokov2019, Ovchinnikov2021a}. $T_N$ decreases with increasing pressure which is consistent with Ref.\,\citenum{Chen2019}.

Next, we present the magnitudes of AHE at low and high field. Fig.\,\ref{fig:AHE} summarizes the magnetotransport data collected at $T=1.5\,$K and $p=1\,$GPa (the 2\,GPa set is similar, see Fig.\,S1). In panel (a) the antisymmetrized Hall resistance is shown for several gate voltages. As already mentioned in relation to Fig.\,\ref{fig:bsymm}, a fully quantized $R_\mathrm{xy}$ can be observed at large $|H|$ at $-18$\,V, the CNP. Away from the CNP, the high-field Hall resistance $R_\mathrm{xy} (-8\,\mathrm{T})$ decays towards zero as plotted in Fig.\,\ref{fig:AHE}(b) by black markers. The fact that its sign is independent of $V_\mathrm{g}$ confirms that the plateau is unrelated to the QHE. The position of the peak in $R_\mathrm{xx}(0\,\mathrm{T})$ vs $V_\mathrm{g}$, which is plotted in the same panel in blue symbols, is consistent with the CNP as expected of a low-field insulator state.

In the inset of Fig.\,\ref{fig:AHE}(a) a zoom of the data at low $|H|$ is plotted, showing the AHE hysteresis loops. The black dashed lines are linear fits of the $-23\,$V Hall signal: their vertical distance characterizes the size of the AHE loop, $R_\mathrm{AH}$ as highlighted by the vertical arrow, while their slope gives the ordinary Hall coefficient (OHE). We have estimated these quantities at several gate voltages and plot them in Fig.\,\ref{fig:AHE}(c). The low-field OHE (blue markers) crosses zero sharply at the CNP while its magnitude is largest near here. This is consistent with a continuous change from hole to electron transport, marking a difference between the trivial insulator state here and the axion insulator state in even-SL MBT, where it has been suggested that the OHE exhibits a plateau at zero as a function of $V_\mathrm{g}$ near the CNP \cite{Liu2020,Li2024}. 

$R_\mathrm{AH}$ (black markers in Fig.\,\ref{fig:AHE}(c)) is largest close to the CNP, and decays fast with $V_\mathrm{g}$ on both sides. It changes sign in the hole regime, which is most apparent in the curves at $2$\,GPa (see the inset of Fig.\,S1(a)). In other words, at this doping, the low-field AHE signal depends on the net magnetization oppositely to the high-field case.

\subsubsection{Temperature-dependence}

\begin{figure}[!t]
    \centering
    \includegraphics[width=\columnwidth]{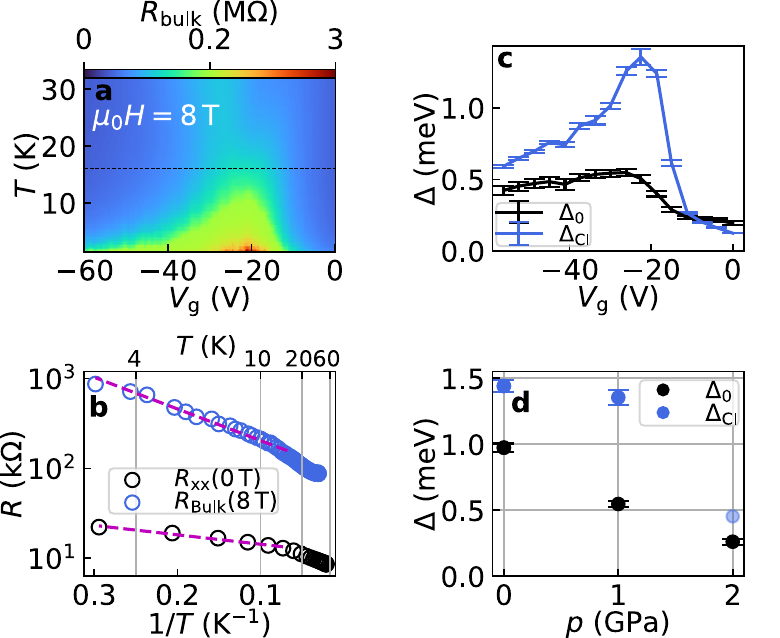}
    \caption{\label{fig:activ} \textbf{Thermal activation in the trivial and Chern insulator states.} (a) The bulk resistance measured at $1\,$GPa and $\mu_0H=8\,$T as a function of gate voltage and temperature. For visibility, the color scale has two linear slopes, with the center at $0.2\,$M$\Omega$. The black dashed line indicates the Néel temperature, $T_N=16\,$K. (b) Arrhenius plot of the 0\,T longitudinal (black) and the 8\,T bulk resistances (blue, from panel (a)) at the CNP. The dashed lines correspond to linear fits of the data below $T_N$. (c) Thermally activated gap sizes below $T_N$ as a function $V_g$ at 1\,GPa. Black and blue markers were obtained from $R_\mathrm{xx}(T)$ and $R_\mathrm{bulk}(T)$ at 0 and 8\,T, and thus correspond to the trivial ($\Delta_0$) and CI ($\Delta_\mathrm{CI}$) gap, respectively. (d) The maximal values of $\Delta_0$ and $\Delta_\mathrm{CI}$ at all pressures. } 
\end{figure}

In order to study the nature of the different insulating states as the system is tuned by hydrostatic pressure, we performed temperature-dependent measurements. In Fig.\,\ref{fig:activ}(a) $R_\mathrm{bulk}$ is shown as a function of $V_\mathrm{g}$ and $T$ in the FM phase at $\mu_0H=8\,$T and $1\,$GPa. It exhibits a maximum value centered near the CNP for all $T$. Moreover, it increases with decreasing temperature as illustrated by the vertical cut in the CNP in Fig.\,\ref{fig:activ}(b), confirming the presence of a gap in the CI state, $\Delta_\mathrm{CI}$. As for the trivial insulator state, we focus on the longitudinal resistance $R_\mathrm{xx}$ at 0\,T. Fig.\,\ref{fig:activ}(b) shows its $T$-dependence in the CNP, which enables estimating the trivial insulator band gap $\Delta_0$. Other measurements for all pressures can be found in the SM. 

The gaps were estimated through Arrhenius analysis at all gates, and the results at 1\,GPa are plotted in Fig.\,\ref{fig:activ}(c): both $\Delta_0$ and $\Delta_\mathrm{CI}$ are maximal close to the CNP. 
The pressure-dependence of their peak values is plotted in Fig.\,\ref{fig:activ}(d). The CI gap is more robust than the trivial gap for all pressures. $\Delta_\mathrm{CI}$ at 2\,GPa is likely underestimated since the FM phase has not yet formed at 8\,T. Rather, we expect that it is comparable to the value at 1\,GPa. The 0 and 1\,GPa points for $\Delta_\mathrm{CI}$ are in the FM phase, and their slowly decreasing trend likely reflects the behavior of the exchange interaction component of the CI gap. In contrast, the trivial gap $\Delta_0$ is significantly suppressed by the pressure. 

\section{Discussion}
\label{sec:disc}

Altogether, our experimental results show the Chern insulating state at high field. This state appears in the ferromagnetic phase, and forms at larger magnetic fields as pressure is increased. Our results also show the presence of a trivial insulator state in the AFM phase instead of the QAHE features expected in odd-SL MBT. Ideally, the parallel magnetization of the top and bottom surface layers opens a sizeable gap in the topological surface states which, according to theoretical calculations, is on the order of $80\,$meV\cite{Deng2020a, Garnica2022}. Hence, in a simple picture, the characteristic energy scale of the QAHE is the exchange energy. The absence of the QAHE might be explained in terms of other energy scales being comparable to the exchange energy, such as the coupling between the topological states at the top and bottom surfaces, although the exact mechanism is up for debate. Other explanations may include that the topological surface state wave function is shifted in real space and the non-trivial gap is effectively reduced \cite{Tan2023, Li2024, Wang2024}, or that the magnetic parameters are different at the surface \cite{Yang2021, Chen2024}.

Another possibility is that the non-trivial exchange gap $\Delta_A$ of the AFM state, already strongly weakened by Mn$_\mathrm{Bi}$ antisite defects as indicated in Ref.\,\citenum{Garnica2022}, is surpassed by a sufficiently strong disorder potential. This would ultimately lead to a trivial insulator phase\cite{Song2016} which has an effective transport gap ($\Delta_0$). Consequently, the intrinsic anomalous Hall conductivity $\sigma_\mathrm{xy}$ is effectively reduced to near zero and strong localization occurs. Upon increasing the magnetic field, transitions occur to the cAFM and then to the FM phase. The FM exchange gap $\Delta_\mathrm{CI}$ might be larger than $\Delta_A$ of the AFM phase, since the exchange field is larger if all layers are aligned (discussed further below) \cite{Deng2020a} or because an external magnetic field may polarize the Mn antisite defects, effectively increasing the magnetization of each layer\cite{Garnica2022}. Hence, $\Delta_\mathrm{CI}$ could dominate the disorder (on the order of $\Delta_0$) and allow Hall quantization, matching the observations. Accordingly, in Fig.\,\ref{fig:bsymm}(a,b) the change in the slope of $R_\mathrm{xy}$ and the drop in $R_\mathrm{xx}$ at the spin-flop (AFM to cAFM) transition at $H_\mathrm{SF}$ indicate the departure from the trivial insulator state and the reappearance of a large intrinsic $\sigma_\mathrm{xy}$.

The anomalous Hall effect in the AFM state is indeed weak and, interestingly, its magnitude $R_\mathrm{AH}$ changes sign when $V_\mathrm{g}$ is tuned as shown in Fig.\,\ref{fig:AHE}(c). This is even more pronounced at 2\,GPa, as shown in Fig.\,S1(b,c). The sign reversal of the AHE in Cr-doped Bi$_2$(Se$_x$Te$_{1-x}$)$_3$ samples has been experimentally observed before and is explained in terms of intrinsic AHE in a disordered system \cite{Zhang2013}. Another explanation for the sign reversal is an extrinsic origin for AHE \cite{Zhang2020}. In any case, the presence of the gate-tunable sign reversal of the AHE supports the possibility of the important role of disorder. 

We have analyzed the magnetic transition fields shown in Fig.\,\ref{fig:bsymm}(d) based on the linear chain model (see SM). The fact that the spin flip (cAFM/FM) transition occurs at much higher fields than the AFM/AFM/cAFM transitions ($H_\mathrm{FM} \gg H_\mathrm{c0}, H_\mathrm{SF}$) suggests that the easy axis anisotropy field $H_a$ in Eq.\ref{eq:linchain_main} is significantly weaker than the antiferromagnetic interlayer exchange $H_E$ (see Fig.\,S7(b)) at all pressures. Moreover, $H_\mathrm{FM}$ increases as a function of pressure while $H_\mathrm{c0}, H_\mathrm{SF}$ remain approximately constant. Based on the model, this is only possible if $H_E$ increases and $H_a$ decreases with pressure (see SM). The former is consistent with expectations \cite{Chen2019, Chong2023a} and can be attributed to the compression along the $c$ axis. The latter may be the result of the reduced distance of nearest-neighbour Mn atoms due to the compression in the $ab$ plane \cite{Yan2019b}.

As for the Néel temperature shown in Fig.\,\ref{fig:bsymm}(c), mean-field considerations predict that $T_N\propto H_F + H_E$ in the bulk limit\cite{Uryu1990,Johnston2017}. Here $H_F$ is the intralayer ferromagnetic coupling, which would appear in a $-\frac{1}{2}H_F \sum_j M_j^2$ term in Eq.\,\ref{eq:linchain_main} and is assumed to be much stronger than the interlayer coupling, $H_F \gg H_E$. The contribution of anisotropy to $T_N$ is negligible in comparison. Therefore the increase in $H_E$ with pressure suggests an even greater decrease in $H_F$. Such an effect has been tied to the intralayer AFM interactions getting stronger due to smaller Mn-Mn distances in the $ab$ plane \cite{Yan2019b,Chen2019}, as it effectively decreases the FM coupling within the SL.

In light of the above, we can discuss the effect of pressure on the theoretically expected band gaps. The exchange energy of the top or bottom surface of the MBT crystal is $\sum_{k} M_k J_{s,k}$ where $J_{s,k}$ is the exchange coupling between the chosen surface and the $k^\mathrm{th}$ layer \cite{Deng2020a}. Assuming only intralayer and nearest-neighbor interlayer exchange and using our previous notations, in the FM or AFM phase this sum is proportional to $H_F \pm H_E $. Neglecting tunnelling between the surfaces or disorder, this predicts that the band gap in the FM phase is $\Delta_\mathrm{CI} \propto H_F + H_E$. Its evolution with pressure based on Arrhenius analysis (see Fig.\,\ref{fig:activ}(d) at 0 and 1\,GPa) qualitatively matches the trend of $T_N$ in Fig.\,\ref{fig:bsymm}(c). As for the AFM gap, with the above assumptions it is predicted to be $\Delta_A \propto H_F - H_E$ and indeed smaller than $\Delta_\mathrm{CI}$. Their difference may be potentially even wider if the layer magnetizations are affected by field-polarizable defects. Therefore, an intermediate disorder potential may surpass $\Delta_A$ but not $\Delta_\mathrm{CI}$

If the disorder is strong enough, localization occurs in the device. In this picture, we expect that as the lattice constants decrease with pressure on the order of a few \% \cite{Chen2019} and atomic wave functions overlap more and more, the localization length $\xi$ increases. Therefore, the average activation energy $\Delta_0$ of hopping transport will decrease \cite{Mott1979, Shklovskii1984, Pollak1991}. According to Anderson localization, the conductivity increases with the localization length: $\sigma_\mathrm{xx}\propto \exp(-L_S/\xi)$, where $L_S$ is the sample size. All these are in good agreement with our experimental findings ($\Delta_0$ in Fig.\,\ref{fig:activ}(d) and $R_\mathrm{xx} \propto \sigma^{-1}_\mathrm{xx}$ in Fig.\,\ref{fig:3x3}), which suggests that pressure weakens the Anderson insulator state by increasing the localization length. At high enough pressure the effect of disorder may become weak enough that it no longer overcomes the AFM gap $\Delta_A$. This would lead to the closure of $\Delta_0$, the reopening of the non-trivial gap and the recovery of the QAHE\cite{Song2016}. However, $\Delta_A$ also decreases with $p$ due to both $H_F$ and $H_E$, which may prevent reaching the QAHE state. 

\section{Summary}
\label{sec:sum}

Our study systematically explores the interplay between pressure, magnetism, and topology in a 5-SL MnBi$_2$Te$_4$ film. At low fields in the AFM phase the near-zero Hall resistance and high longitudinal and bulk resistance indicate a trivial insulator state likely due to disorder. Upon reaching the FM phase by increasing the magnetic field, quantization is recovered. From the analysis of magnetic transition fields based on the magnetoresistance, we estimate that the interlayer exchange coupling $H_\mathrm{E}$ is enhanced by hydrostatic pressure, while the decreasing Néel temperature suggests that the intralayer coupling $H_\mathrm{F}$ is reduced. Although the effective trivial band gap $\Delta_0$ was also reduced, the quantum anomalous Hall effect was not recovered under applied pressure, suggesting that 2\,GPa was insufficient to fully overcome extrinsic factors such as disorder in our device.

These findings provide new insights into pressure-driven phase transitions in magnetic topological materials. Further studies in material engineering and heterostructure fabrication may be required to stabilize and enhance the QAHE in MBT-based systems.

\section{Methods}
\label{sec:methods}

MBT crystals were grown by the self-flux method at UCLA \cite{Ni2024}. We obtained the thin flakes using standard mechanical exfoliation onto a Si wafer covered with 280-nm-thick SiO$_2$. We determined the flake thickness by its optical contrast. In order to protect its surface from contamination during the nanolithography process used for fabricating Ohmic contacts, we utilized a shadow mask evaporation technique. The contacted MBT flake was then covered with a thin flake of hexagonal boron nitride (hBN). The whole device fabrication process was done in an argon-filled glovebox, so the air sensitive MBT was protected throughout all steps. 

The PCB chip carrier method \cite{Fulop2021} enables mechanical bonding and a relatively easy and rapid exchange of samples in the pressure cell, compared to directly fixing the chip with epoxy and manually bonding the floating wires. Therefore, the prerequisites for samples (regarding the size and spacing of bonding pads) are less rigorous. Essentially, even samples that were not specifically designed for pressure cell measurements are viable if the lateral size of the chip does not exceed 3\,mm. 

 The (anti)symmetrized longitudinal (Hall) data in Figures\,\ref{fig:bsymm},\,\ref{fig:AHE} were calculated following
\begin{align}
\label{eq:antisymm1}
R_\mathrm{xx(S)}^{\uparrow \downarrow} \left( H \right) = \left[ R_\mathrm{xx}^\uparrow \left( \pm H \right) + R_\mathrm{xx}^\downarrow \left( \mp H \right) \right] / 2, \\
\label{eq:antisymm2}
R_\mathrm{xy(AS)}^{\uparrow \downarrow} \left( H \right) = \pm \left[ R_\mathrm{xy}^\uparrow \left( \pm H \right) - R_\mathrm{xy}^\downarrow \left( \mp H \right) \right] / 2,
\end{align}
where $\uparrow$ and $\downarrow$ refer to the direction of the field sweep. 

\section{Acknowledgements}
\label{sec:ack}

This research was supported by the Ministry of Culture and Innovation and the National Research, Development and Innovation Office within the Quantum Information National Laboratory of Hungary (Grant No.\ 2022-2.1.1-NL-2022-00004 and UNKP-23-4-I-BME-36), by OTKA grants No.\ K138433 and K134437 and the NKKP STARTING grant No.\ 150232. We acknowledge funding from the Multi-Spin and 2DSOTECH FlagERA networks, the 2DSPIN-TECH Flagship project, the Alexander von Humboldt Foundation, the European Research Council ERC project Twistrain, the TRILMAX Horizon Europe consortium (Grant No. 101159646), and COST Action CA 21144 superQUMAP. The work in Xu group is supported by the Air Force Office of Scientific Research (AFOSR) grant FA9550- 626 23-1-0040. S.-Y.X. acknowledges the Sloan foundation and Corning Fund for Faculty Development. K.W. and T.T. acknowledge support from the JSPS KAKENHI (Grant Numbers 20H00354 and 23H02052) and World Premier International Research Center Initiative (WPI), MEXT, Japan.

Device fabrication was performed by Y.-F.L., A.G. and T.H. with the guidance of S.-Y.X. Measurements were performed by A.M. Experimental evaluation was done by A.M., theoretical evaluation by E.T. and O.L. A.M. and E.T. wrote the paper and all authors discussed the results and worked on the manuscript. The project was guided by S.-Y.X., E.T., S.C. and P.M.

\FloatBarrier 
\bibliography{MBT}

\balancecolsandclearpage


\balancecolsandclearpage
\onecolumngrid
\appendix*

\counterwithin*{figure}{part}
\stepcounter{part}
\renewcommand{\thefigure}{S\arabic{figure}}

\counterwithin*{equation}{part}
\stepcounter{part}
\renewcommand{\theequation}{S\arabic{equation}}

\section*{Supplementary Material}

\subsection{Additional experimental data}
\label{subsec:extras}

In this section we present further details of our measurements.

\subparagraph{AHE data at 2\,GPa:}

\begin{figure}[tbh]
    \centering
    \includegraphics[width=0.65\columnwidth]{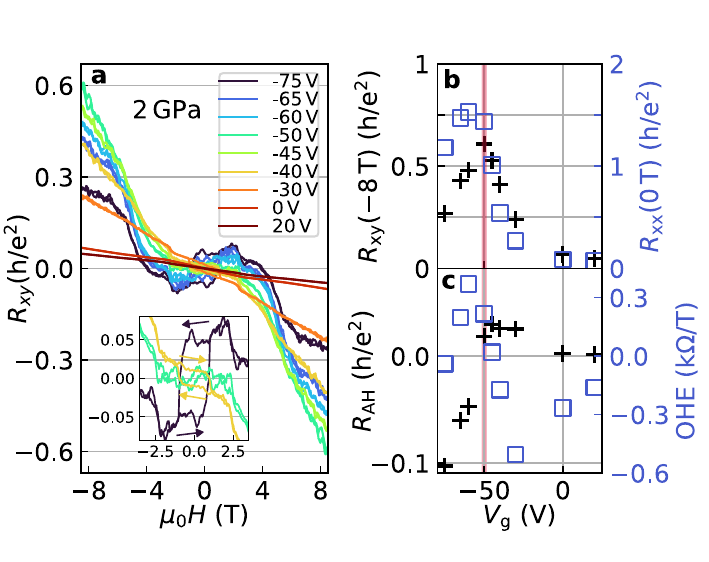}
    \caption{\label{fig:AHE_2gpa} \textbf{AHE at 2 GPa and 1.5K.} \textbf{(a)} Antisymmetrized $R_{\mathrm{xy}}$ as a function of $H$ at a series of gate voltages. Inset: a zoom of the data. \textbf{(b)} $R_{\mathrm{xy}}$ at -8\,T (black markers) and $R_{\mathrm{xx}}$ at 0\,T (blue) as a function of $V_g$. The CNP is marked by a red line. \textbf{(c)} The size $R_\mathrm{AH}$ of the hysteresis loop (black symbols) and the ordinary Hall coefficient (OHE, in blue) at low field.}
\end{figure}

In Fig.\,\ref{fig:AHE_2gpa} the antisymmetrized Hall data at 2\,GPa and 1.5\,K and its analysis is shown, similarly to Fig.\,4 in the main text. Panel (a) shows the anomalous Hall signal at a series of gate voltages. The CI state at high field was not observed, although presumably it appears at higher fields. The hysteresis at low field is still present, and its sign change is especially visible as illustrated in the inset. Fig.\,\ref{fig:AHE_2gpa}(b) shows that the Hall resistance at -8\,T (black markers) peaks in the vicinity of the CNP, as does the zero-field longitudinal resistance (blue). Panel (c) shows the size of the low-field hysteresis $R_\mathrm{AH}$ versus the gate voltage (black), and the slope of the linear component of the Hall resistance, the ordinary Hall coefficient OHE (blue). The results are qualitatively similar to those at 1\,GPa.

\begin{figure}[!h]
    \centering
    \includegraphics[width=1\columnwidth]{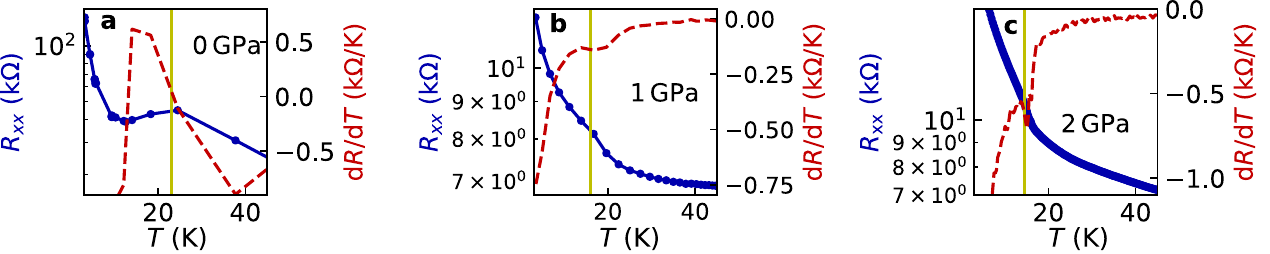}
    \caption{\label{fig:bumpsOnRT} Four-probe resistances and their derivatives (red dashed lines) as a function of temperature for a) $0\,$GPa, b) $1\,$GPa and for c) $2\,$GPa pressure. The antiferromagnetic phase transition manifests as a small resistance peak as a function of $T$ due to spin scattering mechanisms. The value of $T_\mathrm{N}$ may also be obtained by following the derivative of the resistance, as highlighted by vertical lines.}
\end{figure}

\subparagraph{Estimating the Néel temperature:}

The temperature-dependence of the four-probe resistance at zero field is shown in Fig.\,\ref{fig:bumpsOnRT}. Around the Néel temperature $T_\mathrm{N}$ there are intensified spin fluctuations which add a peak to the expected trend of $R_\mathrm{\mathrm{xx}}(T)$ \cite{Lee2019}, which helped determine the Néel temperature plotted in Fig.\,3(c). While this peak is apparent at 0\,GPa (Fig.\,\ref{fig:bumpsOnRT}(a)) in resistances, at higher pressure (b-c) the sloping background makes its identification difficult. Therefore, here we estimated $T_N$ from the center of the $\sim$ shape of the derivatives $dR/dT$ shown by red dashed lines.

\begin{figure}[!t]
    \centering
    \includegraphics[width=0.75\columnwidth]{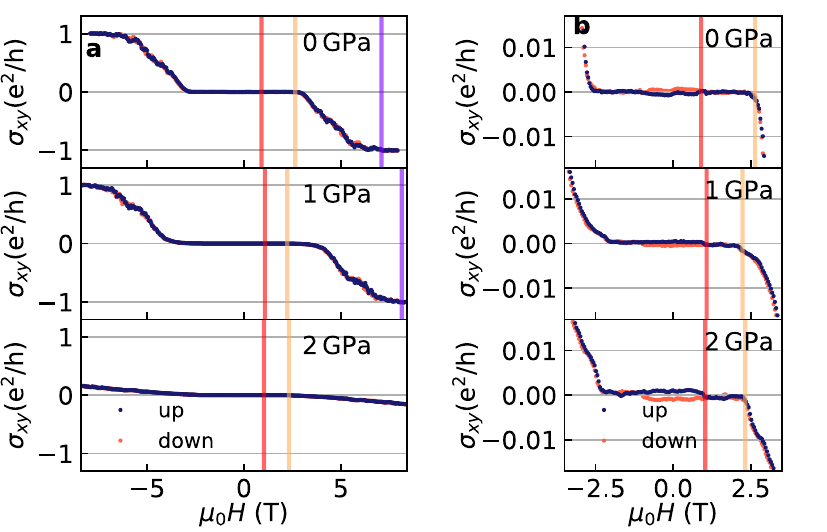}    
    \caption{\label{fig:bsymmSM} (a) Up (black) and down (red) field-sweeps of the antisymmetrized Hall conductivity close to the CNP at different pressures. (b) Zoom-in to panel (a). Red lines represent the edge of the hysteresis loop ($H_\mathrm{c0}$), orange lines the drop in $\sigma_\mathrm{xy}$ and the edge of the AFM phase ($H_\mathrm{SF}$), and purple lines the onset of FM order ($H_\mathrm{FM}$). }
\end{figure}

\subparagraph{Extracting the magnetic transition fields:}

In Fig.\,\ref{fig:bsymmSM} we show magnetoconductance curves measured near the CNP at all pressures. While the Hall voltage signal did not diverge at 1 and 2\,GPa in the AFM state, enabling the calculation of resistance, it was significantly affected by the mixing of the divergent longitudinal signal at 0\,GPa as discussed in the main text. This made the extraction of the two lower magnetic transition fields, $H_\mathrm{c0}$ for AFM-AFM and $H_\mathrm{SF}$ for AFM-cAFM (spin flop), difficult. Therefore we estimated them by calculating the Hall conductivity $\sigma_\mathrm{xy}$, which is plotted in the top panel of Fig.\,\ref{fig:bsymmSM} (a) in the full field range, and in the top panel of (b) near zero field. The black and red lines show the up and down sweeps of the field, demonstrating the presence of hysteresis and enabling the extraction of the transition fields. For comparison, $\sigma_\mathrm{xy}$ is plotted at 1, 2\,GPa as well.

\subparagraph{Maps of temperature and gate-dependence:}

\begin{figure}[tbh]
    \centering
    \includegraphics[width=0.75\columnwidth]{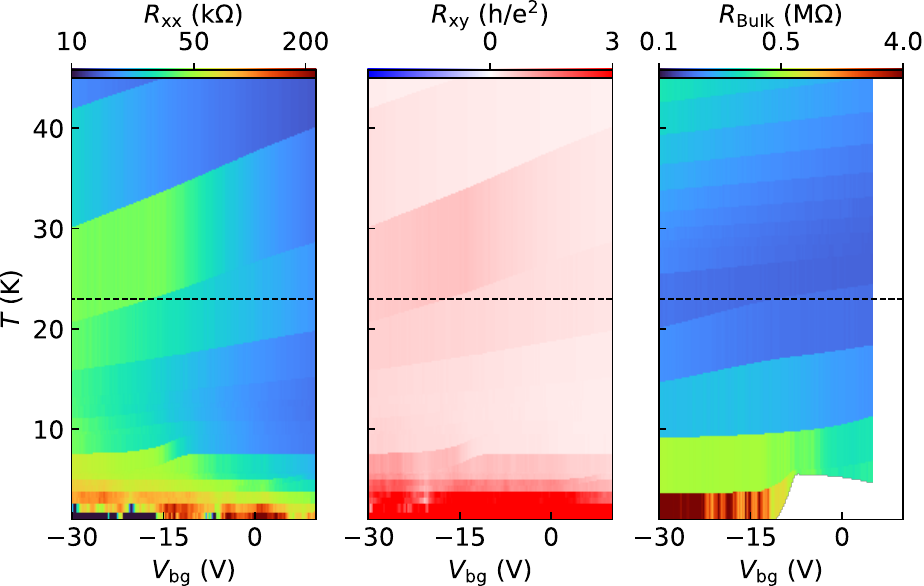}
    \caption{\label{fig:act0gpa} Temperature-dependent resistance measurements as a function of gate voltage at 0\,GPa. $R_\mathrm{xx/xy}$ were measured at 0\,T, $R_\mathrm{bulk}$ at 8\,T. The Néel temperature is indicated by a black dashed line.}
\end{figure}

\begin{figure}[tbh]
    \centering
    \includegraphics[width=0.75\columnwidth]{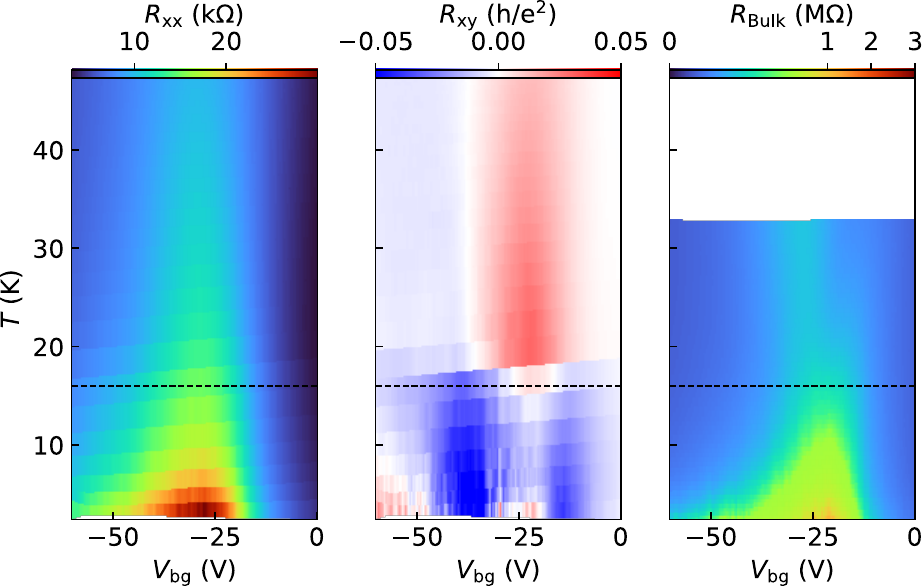}
    \caption{\label{fig:act1gpa} Temperature-dependent resistance measurements as a function of gate voltage at 1\,GPa. $R_\mathrm{xx/xy}$ were measured at 0\,T, $R_\mathrm{bulk}$ at 8\,T. The Néel temperature is indicated by a black dashed line.}
\end{figure}

\begin{figure}[tbh]
    \centering
    \includegraphics[width=0.75\columnwidth]{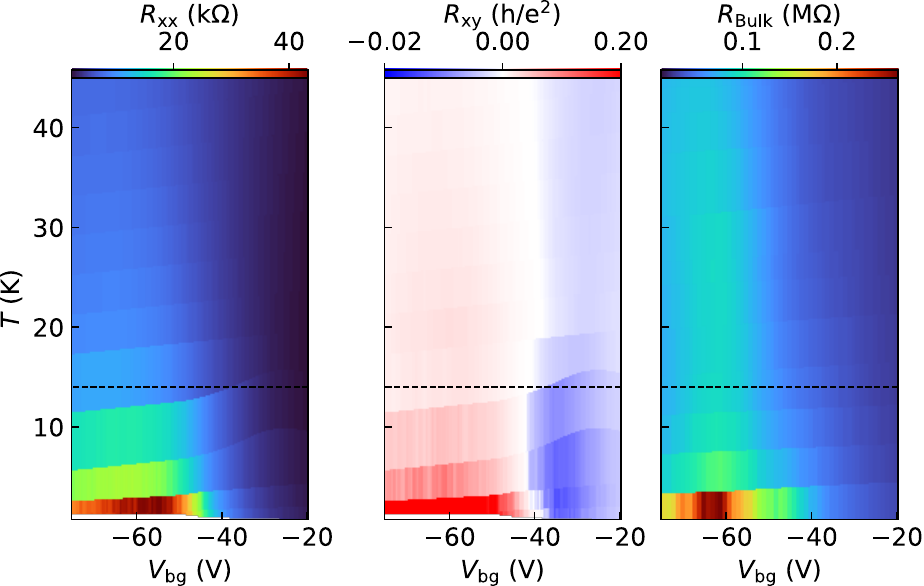}
    \caption{\label{fig:act2gpa} Temperature-dependent resistance measurements as a function of gate voltage at 2\,GPa. $R_\mathrm{xx/xy}$ were measured at 0\,T, $R_\mathrm{bulk}$ at 8\,T. The Néel temperature is indicated by a black dashed line.}
\end{figure}

Figures\,\ref{fig:act0gpa},\,\ref{fig:act1gpa},\,\ref{fig:act2gpa} show resistance maps at 0, 1 and 2 GPa, respectively. The longitudinal and Hall resistances as a function of gate voltage and temperature are plotted at 0\,T, while $R_\mathrm{bulk}$ is plotted at 8\,T. Below $T_N$ these measurements illustrate the trivial and Chern insulator states without and with magnetic field, respectively. The zero-field $R_\mathrm{xy}$ at 1\,GPa (Fig.\,\ref{fig:act1gpa}) clearly shows the disappearance of the AHE in the vicinity of $T_N$. However, since it carries a component from $R_\mathrm{xx}$, this may be hidden, as is the case for 0 and 2\,GPa. 

\FloatBarrier

\subsection{The linear chain model}
\label{subsec:linchain}

An effective classical model similar to the Stoner-Wohlfarth model can be used to describe the magnetic phase transitions in an A-type AFM like MBT. In this section we discuss the model and the magnetic phases and transition fields predicted by it. In the next section we will use it to estimate the interlayer AFM coupling $H_E$ and the anisotropy $H_a$ and demonstrate their pressure-dependence.

In MBT the intralayer FM coupling is relatively strong, which enables describing each layer with a macrospin magnetization, saturated in the low-temperature limit. We assume an easy-axis anisotropy along the $z$ axis (out of plane for MBT) and set a magnetic field parallel to it. Therefore we need only consider the orientation of each layer, and the energy function can be simplified to the following form \cite{Deng2020a, Ovchinnikov2021a, Yang2021, Bac2022, Chen2024}:
\begin{equation}
    \label{eq:linchain}
    f = H_E \sum_{j=2}^N \cos \left( \varphi_j - \varphi_{j-1} \right) - \frac{H_a}{2} \sum_{j=1}^N \cos ^2 \varphi_j 
    - H \sum_{j=1}^N \cos \varphi_j.
\end{equation}
Here $\varphi_j \in [0,2\pi[$ is the magnetization orientation of the $j^{th}$ layer relative to $z$, $N$ is the layer number, $H_E > 0$ is the interlayer AFM exchange, $H_a >0$ is the anisotropy energy selecting $z$ ($\varphi = 0$ or $\pi$) as the easy axis, and the final term is the Zeeman term. 

\vspace{0.2cm}

\subsubsection{Analytical solutions}

\vspace{0.2cm}

To determine the borders of the magnetic phases, we need to consider the gradient of $f$, as well as the Hessian matrix with elements $A_{jk} = \partial^2_{\varphi_j,\varphi_k} f $. $A$ is a tridiagonal matrix where the super, sub, and main diagonals are, respectively,
\begin{align}
    \label{eq:Hessian}
    A_{j,j+1} &= H_E \left( 1 - \delta_{jN} \right) \cos \left( \varphi_{j+1} - \varphi_j \right) = b_j, \\
    A_{j,j-1} &= H_E \left( 1 - \delta_{j1} \right) \cos \left( \varphi_{j} - \varphi_{j-1} \right) = d_j, \\
    A_{j,j} &= - \left( b_j + d_j \right) + H_a \cos 2 \varphi_j + H \cos \varphi_j = a_j
\end{align}
where $\delta_{jk}$ is the Kronecker-delta.
\begin{equation}    
    A = \begin{pmatrix}
        a_1 & b_1                                \\
        d_1 & a_2  & b_2                         \\
            & d_2  & a_3   & \ddots              \\
            &      &\ddots & \ddots  & b_{N-1}   \\
          &        &       & d_{N-1} & a_N
      \end{pmatrix}.
\end{equation}
We will use the eigenvalues of this matrix in the various magnetic phases that satisfy $\bm{\nabla}f=\bm{0}$ to determine the field regimes where they provide local minima in the energy. 

We shall denote the phases by the same acronyms as before, but specify their net magnetization by a subscript. For instance, FM$_{+N}$ will mean the fully polarized state where all layer magnetizations are aligned with a positive field $H$ ($\uparrow \uparrow \uparrow \uparrow \uparrow$ for $N=5$). The net dimensionless magnetization, defined as
\begin{equation}
    M = \sum_{j=1}^N \cos \varphi_j,
\end{equation}
is $M = +N$ in this phase (the saturation magnetization of a SL is taken as 1).

\subparagraph{AFM-like phases:} \mbox{}
\vspace{0.1cm}

For an even number of layers we label the antiferromagnetic state near $H=0$ as AFM$_{0}$, since $M=0$ (for $N=4$, it is $\downarrow \uparrow \downarrow \uparrow$ or its reverse). For odd $N$, the states AFM$_{\pm 1}$ compete at low $|H|$: AFM$_{+1}$ is where $(N + (-) 1)/2$ layer magnetizations are (anti)parallel with $H$, and AFM$_{-1}$ is the reverse. For $N=5$ they are $\uparrow \downarrow \uparrow \downarrow \uparrow$ and $\downarrow \uparrow \downarrow \uparrow \downarrow$. We will also use the notation AFM for AFM-like states where SL magnetizations are along $\pm z$ but not fully polarized ($|M|$ is an integer and $<N$).

\subparagraph{cAFM and FM:} \mbox{}
\vspace{0.1cm}

For weak anisotropy the canted AFM (cAFM) phase is possible. In this state the magnetic moments are not parallel with the easy axis or the applied field, and due to the AFM coupling, their orientation alternates from layer to layer, like \rotatebox[origin=c]{22}{$\uparrow$}\rotatebox[origin=c]{-45}{$\uparrow$}\rotatebox[origin=c]{45}{$\uparrow$}\rotatebox[origin=c]{-45}{$\uparrow$}\rotatebox[origin=c]{22}{$\uparrow$} for $H>0$ and $N=5$. If we start increasing $H$ from the cAFM$_+$ phase ($+$ indicates $M>0$), the system eventually reaches the fully polarized FM$_{+N}$ state ($\uparrow \uparrow \uparrow \uparrow \uparrow$). The transition to FM$_{+N}$ is second-order, the cAFM-FM transition or its reverse does not produce a hysteresis. This saturation field $H_\mathrm{FM}$ is the edge of stability of the cAFM$_{+}$ state when increasing $H$. It is also the edge of stability of the FM$_{+N}$ state when decreasing $H$. In general, the edge of stability means that here the corresponding local minimum in $f$ disappears while changing $H$ and the system must find another. In this example (cAFM$_+$ to FM$_{+N}$ or vice versa), the local minimum does not disappear and there is a seamless (hence second-order) transition. A magnetization curve simulated for $N=5$ is illustrated in Fig.\,\ref{fig:linchain}(f) by a black line (simulations are discussed further below), the saturation at $H_\mathrm{FM}$ is indicated by the purple line.

$H_\mathrm{FM}$ can be determined by calculating the (positive) eigenvalues of $A$ in the FM$_{+N}$ state, and find where the smallest of them reaches zero when decreasing $H$. Since $\varphi_j = 0$, the matrix elements are $b_j = d_j = d = H_E$, $a_{j=2:N-1} = a = - 2 H_E + H_a + H$ while $a_{j=1,N} = a+d = - H_E + H_a + H$ due to only a single neighboring layer. Using trial eigenvectors with $k^{th}$ components $u_k = e^{\gamma k} + c \cdot e^{-\gamma k}$, the solutions are of the form $\gamma_m = \left( m-1 \right) \pi i / N$ where $i$ is the imaginary unit, and the eigenvalues are $\lambda_m = H_a + H - 2 H_E \left( 1 - \cos \frac{\left( m-1 \right) \pi}{N} \right) $. The smallest is $\lambda_N$, therefore enforcing $\lambda_N = 0$ gives the edge of stability of the FM$_{+N}$ state, which is also the border of the cAFM$_+$ phase. As stated above, the field where it occurs is the saturation field
\begin{align}
    \label{eq:sat}
    H_\mathrm{FM} &= H_\mathrm{FM0} - H_a, \mathrm{where} \\
    H_\mathrm{FM0} &= 4 H_E \cos^2 \frac{\pi}{2 N}.
\end{align}
Here $H_\mathrm{FM0}$ is the saturation field at $H_a=0$, which is consistent with Ref.\,\citenum{Wang2019}. The saturation field for the opposite, FM$_{-N}$ state is $-H_\mathrm{FM}$: the edge of stability for this state when increasing $H$, and the edge of stability for the cAFM$_-$ state when decreasing $H$. This is visible in Fig.\,\ref{fig:linchain}(f) in the symmetry of the saturated regions of the simulated magnetization (black line). These borders are plotted as diagonal purple lines in Fig.\,\ref{fig:linchain}(a) for not too high anisotropy.

As for stronger anisotropy, above a certain $H_a$ the cAFM phase is no longer a \textit{global} energy minimum for any $H$. Here we determine this critical value of $H_a$. In this regime the global minimum of the AFM$_{0}$ or the AFM$_{+1}$ state (depending on the parity of N) becomes equienergetic with that of the FM$_{+N}$ state at a field we label as $H_\mathrm{AFM}$, which is $H_E$ for $N=2$ layers and $2H_E$ for $N>2$. The critical $H_a^c$ above which the cAFM state cannot be a global minimum is determined by the equation $H_\mathrm{FM} (H_a^c) = H_\mathrm{AFM}$, producing
\begin{equation}
    H_a^c = \Biggl\{ \begin{matrix} H_E & \mathrm{if}~N=2, \\
    2H_E \left( 2 \cos ^2 \frac{\pi}{2N} - 1 \right) & \mathrm{if}~N>2. \end{matrix}
\end{equation}
Above $H_a^c$ the border of the FM phase being a global minimum is $H_\mathrm{AFM}$, independent of $H_a$, which is indicated by the purple lines turning vertical in Fig.\,\ref{fig:linchain}(a). 

\vspace{0.2cm}

\subsubsection{Numerical calculations}

\vspace{0.2cm}

\subparagraph{Looking for the global energy minimum:} \mbox{}
\vspace{0.1cm}

\begin{figure}[t!]
\includegraphics[width=0.8\columnwidth]{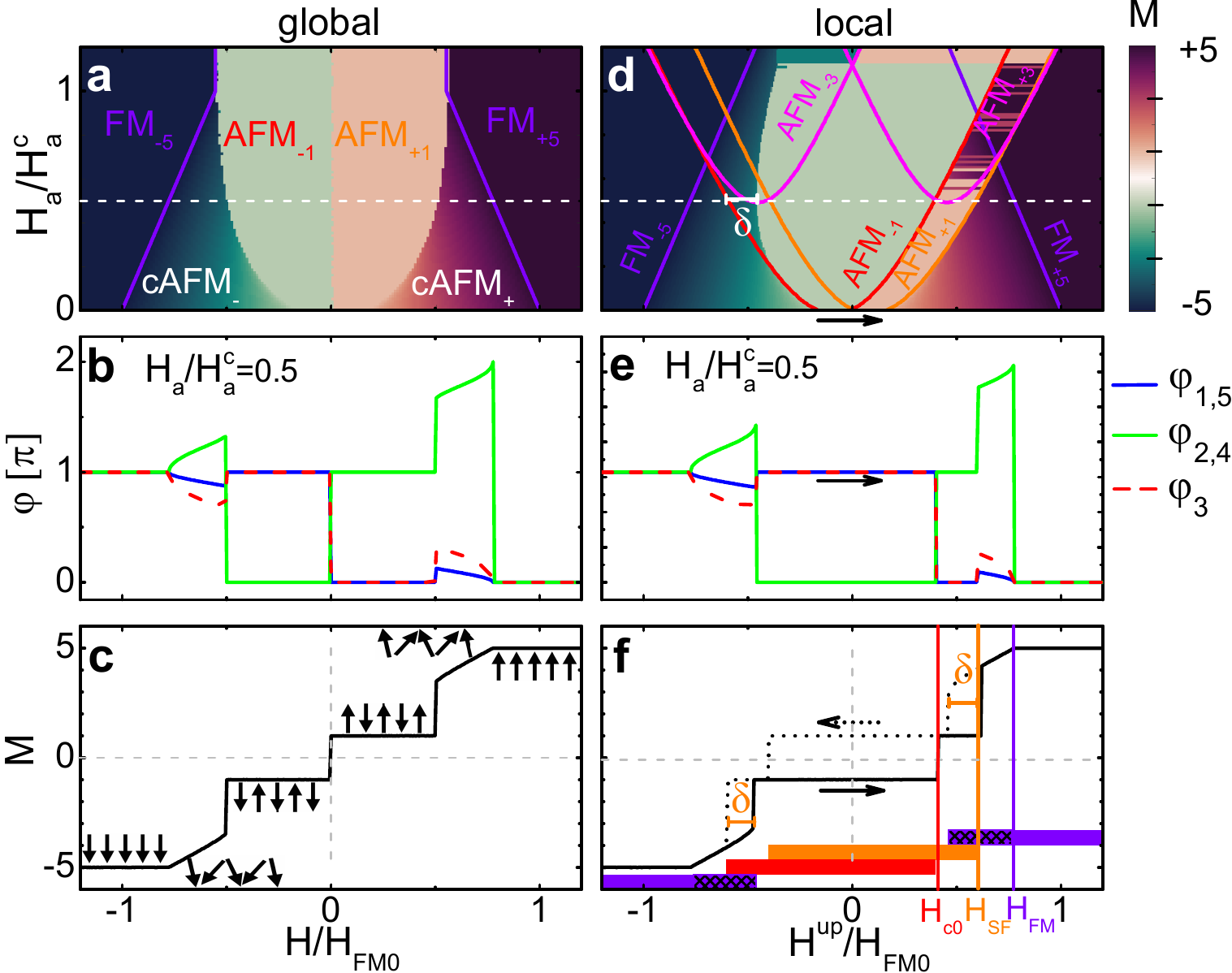}
\caption{\label{fig:linchain} Magnetic simulations for $N=5$. (a) Numerical global energy minimum search result showing a colormap of $M / M_\mathrm{sat}$ as a function of the dimensionless field $H / H_\mathrm{FM0}$ and anisotropy $H_a / H_a^c$. The purple lines are the borders of the FM phases. (b) The angles $\varphi_j$ of individual layer magnetizations, and (c) the net easy axis magnetization $M / M_\mathrm{sat}$ from a global minimum search, along the white dashed line ($H_a / H_a^c = 0.5$) in (a). (d) Numerical local minimum search result while increasing $H$ (upsweep). Lines show the analytical stability borders of various phases based on the Hessian $A$. (e) The angles and (f) the net magnetization during an upsweep at $H_a / H_a^c = 0.5$. The latter also shows the corresponding downsweep as a dotted line. The colored rectangles highlight the stability regions of the states: purple for FM, cross-hatched purple for cAFM, orange and red for AFM$_{\pm 1}$.}
\end{figure}

We have numerically simulated the possible magnetic configurations. We performed a global minimum search in the phase space $\{\varphi_j\}$ for several values of $H$ and $H_a$ by using a brute force approach combined with the gradient method, optimized to the problem by exploiting the symmetries of Eq.\,\ref{eq:linchain}. The results are shown in Fig.\,\ref{fig:linchain}(a) by plotting the magnetization $M$ in a colormap as a function of $H/H_\mathrm{FM0}$ and $H_a/H_a^c$. The numerical approaches match well the analytical calculations: the cAFM state is indeed a global minimum between the purple lines up to $H_a = H_a^c$, the FM states occur outside these lines, while at low $H$ the AFM states are the most stable. It also gives the AFM-cAFM boundaries according to global minimum considerations, which is not possible analytically. An example of the solutions along a horizontal cut at $H_a / H_a^c = 0.5$ are shown in Fig.\,\ref{fig:linchain}(b,c). The AFM$_{\pm 1}$ states with $M = \pm 1 $ are easily recognizable. Panel (b) demonstrates that in the cAFM phase the angles $\varphi_j$ of next nearest neighbor layers (1-3, 2-4 or 3-5), while similar, are not equal. Therefore a two-sublattice model would be inadequate to describe a system with such a low number of layers. We note that for an even number of layers and $N>2$, the global minimum map is similar but has three low-$H$ regions instead of one: for states AFM$_{-2,0,+2}$ from left to right, except for very low $H_a$ where only a narrow AFM$_0$ region may exist (not shown).

\subparagraph{Looking for local energy minima:} \mbox{}
\vspace{0.1cm}

In order to get a better picture about possible hystereses, first we analytically calculated the stability boundaries (using the eigenvalues of the Hessian $A$) of the AFM and the AFM-like states which possess $M = \pm 1 , \pm 3$. These were determined by finding $\nabla_\varphi f = 0$ as well as $A$ being a positive definite matrix. The results are plotted as colored lines in Fig.\,\ref{fig:linchain}(d): orange and red for the AFM$_{\pm1}$ states, and magenta for the AFM-like states AFM$_{\pm 3}$ which have the configurations $\uparrow \uparrow \downarrow \uparrow \uparrow$ and $\downarrow \downarrow \uparrow \downarrow \downarrow$. Within these regions (above the lines) the states are local minima in the energy landscape. We note that the configurations $\downarrow \uparrow \uparrow \uparrow \uparrow$ and $\uparrow \downarrow \uparrow \uparrow \uparrow$ (and similar ones with reversed spins or reversed order) also produce $M = 3$ (or $-3$) but are contained within the regions outlined in magenta. The border between the cAFM and FM states is $\pm H_\mathrm{FM}$, plotted by purple lines. The FM states are stable for $H$ outside these lines, while cAFM states can be stable between them, although not everywhere.

Besides the above analytical calculations, we performed numerical simulations at a series of fields $H$ at different $H_a$ to find the local energy minima in $\{\varphi_j\}$ space. First we selected the value of $H_a$ and the $H$ sweep direction, then set the starting point of $H$ below $- H_\mathrm{FM}$ or above $+ H_\mathrm{FM}$, depending on the direction. The starting state $\{ \varphi_j \}$ was the corresponding FM state, for example FM$_{-N}$ for an upsweep. After each incremental change in $H$, the gradient and the Hessian were checked to see if the latest energy minimum became unstable. If it did, a random walk was performed until the gradient method could converge to another local minimum. The results for an upsweep of $H$ are plotted in Fig.\,\ref{fig:linchain}(d) by a colormap of $M$, and the specific solution at $H_a / H_a^c = 0.5$ is shown in panels (e,f). In Fig.\,\ref{fig:linchain}(f) the downsweep is also plotted, demonstrating hysteresis loops at the AFM-AFM and AFM-cAFM transitions, the latter's size denoted by $\delta$ in panels (d,f). 

Let us follow $M$ during the upsweep along Fig.\,\ref{fig:linchain}(d) and (f): the right edge of the FM$_{-5}$ state in the simulation matches the analytical $-H_\mathrm{FM}$ (purple line in (d)) without a hysteresis loop since the FM-cAFM transition is second-order. Next we arrive at the right edge of the cAFM$_-$ state at an analytically unknown field and switch to the only possible minimum, AFM$_{-1}$. Its right edge is correctly predicted from the Hessian (red lines in (d,f) denoted by $H_\mathrm{c0}$) and the system switches to the AFM$_{+1}$ state. The right edge of this is also correctly predicted (orange lines, $H_\mathrm{SF}$). The following cAFM$_+$ state seamlessly turns into FM$_{+5}$ at $+H_\mathrm{FM}$ as expected (purple again). The stability regions of the phases based on the calculations are also highlighted in the bottom of Fig.\,\ref{fig:linchain}(f) by colored rectangles. 

At higher $H_a$, when a state destabilizes upon increasing $H$, multiple local minima may be possible. As a result the AFM$_{\pm 3}$ states can also be randomly observed in the numerical simulation (Fig.\,\ref{fig:linchain}(d)) within the predicted regions, with their right edges matching the calculated curves (magenta lines). The downsweep map may be produced without running a simulation: it is given by $M^\mathrm{down} (H) = - M^\mathrm{up} (-H) $, which is illustrated by an actual calculated downsweep curve in Fig.\,\ref{fig:linchain}(f) (dotted line). The exception is when there are multiple available local minima which introduce randomness into the picture. Based on Fig.\,\ref{fig:linchain}(d) this may occur above $H_a / H_a^c > 0.5$.

\subsection{Transitions in the linear chain model}
\label{subsec:linchain_fit}

Here we discuss the experimentally observed phase transitions in light of the linear chain model. When plotting the transition fields, we always follow the same colors as in Fig.\,3 in the main text, see for example Fig.\,\ref{fig:linchain}(f).

Regarding the AFM$_{\pm1}$ phases, a hysteresis and a finite switching field $H_\mathrm{c0}$ between them are expected. They are, indeed, observed in experiment. However, at the start of the cAFM phase (spin-flop field $H_\mathrm{SF}$, orange lines in Fig.\,3 and Fig.\,\ref{fig:linchain}(f)) we expect a sudden jump in $R_\mathrm{xy}$ due to the abrupt change in the net magnetization, but only observe a change in its slope. This discrepancy may be due to layer-dependent magnetic parameters (potentially from defects), the presence of domains \cite{Bac2022}, or that here $R_\mathrm{xy}$ is likely dominated by an intrinsic AHE which is not exactly proportional to $M$. Moreover, based on Fig.\,\ref{fig:linchain}(f), we expect different $H_\mathrm{SF}$ values between up- and downsweeps of $H$, i.e. a hysteresis between the AFM and cAFM phases. In contrast, the magnetoresistance curves lack a hysteresis, but we should notice in Fig.\,\ref{fig:linchain}(d) that the expected width $\delta$ of this hysteresis loop becomes negligible below $H_a / H_a^c \approx 0.25$. As we shall see, the parameters truly fall in this range.

\subparagraph{Qualitative evaluation of $H_E, H_a$ versus pressure:} \mbox{}
\vspace{0.1cm}

\begin{figure}[h]
\includegraphics[width=1\columnwidth]{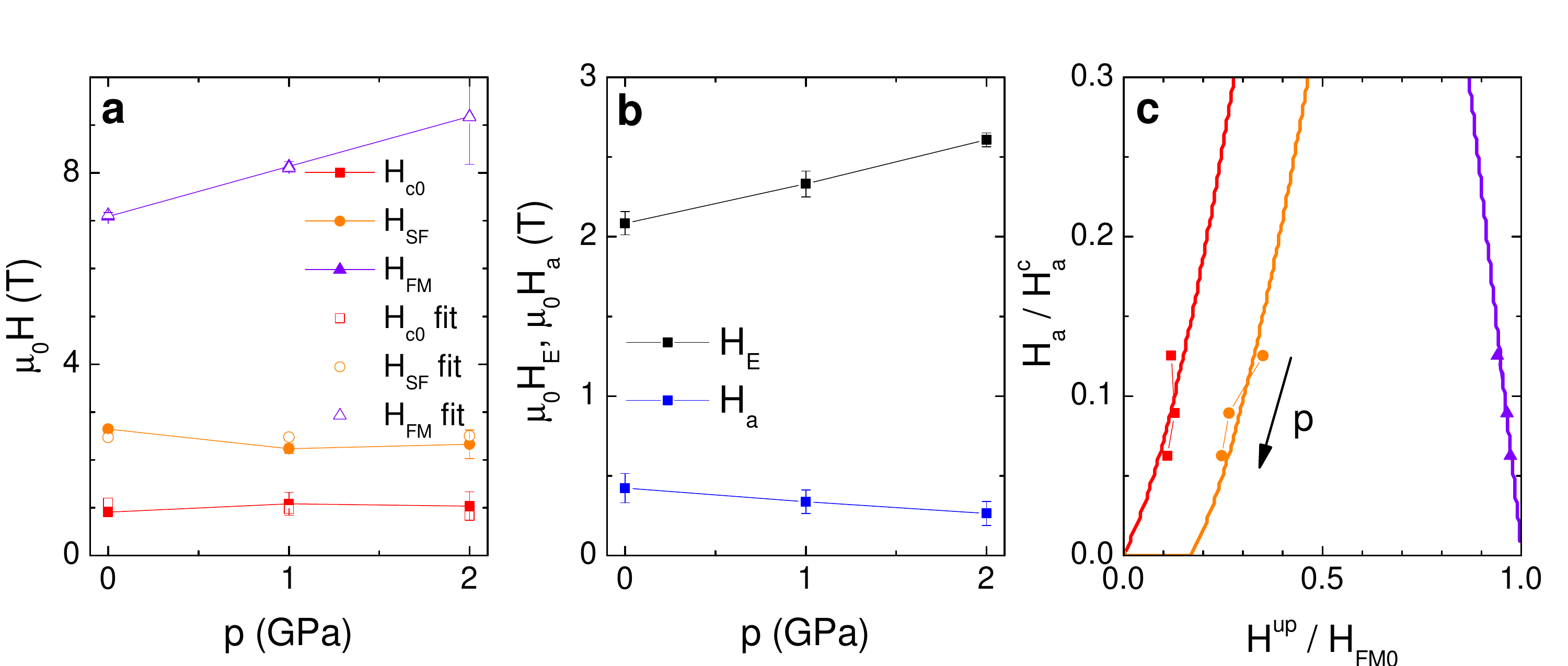}
\caption{\label{fig:linchain_fit} (a) The experimentally extracted transition fields vs pressure (solid symbols) and their fits (empty symbols). (b) The fit parameters $H_E$ and $H_a$ versus pressure. (c) The transition fields (solid symbols) rescaled by $H_\mathrm{FM0}$ on the analytical phase diagram. The solid lines are the same as in Fig.\,\ref{fig:linchain}(d). The arrow indicates that $H_a / H_a^c$ decreases with pressure.}
\end{figure}

First, we explain the qualitative assessment in the main text about the change of magnetic parameters $H_E, H_a$ with increasing pressure. Fig.\,\ref{fig:linchain_fit}(a) shows in solid symbols that the experimental $H_\mathrm{FM}$ increases with pressure, while $H_\mathrm{c0}, H_\mathrm{SF}$ remain approximately constant. Here the value for $H_\mathrm{FM}$ at 2\,GPa was linearly extrapolated, from its values at 0 and 1\,GPa, to be $\sim 9.2$\,T. It is likely underestimated, since when we estimate it another way, by linearly extrapolating $R_\mathrm{xy} (p=2\,\mathrm{GPa})$ in Fig.\,3(a) from the range $6-8.5$\,T and look for the field where $-1\,e^2/h$ is likely reached, we get approximately 16\,T. 

Based on Eq.\,\ref{eq:sat}, $H_\mathrm{FM}$ increasing with pressure is only possible if $H_E$ increases enough compared to the change in $H_a$, meaning 
\begin{equation}
\label{eq:dHE1}
    \delta H_E \cdot 4 \cos^2 (\pi / 2 N) > \delta H_a.
\end{equation}
In contrast, $H_\mathrm{c0}$ and $H_\mathrm{SF}$ attributed to the edges of the AFM$_{\pm1}$ states for $H>0$ are expected to increase monotonously with increasing $H_a / H_a^c$ as shown in Fig.\,\ref{fig:linchain}(d). Approximating either transition field linearly on the phase diagram, $H_\mathrm{c0(SF)} / H_\mathrm{FM0} \approx a + b \cdot H_a / H_a^c$ where $a,b>0$. Consequently, $H_\mathrm{c0(SF)} \approx a\cdot H_\mathrm{FM0} + b \cdot H_a \cdot H_\mathrm{FM0} / H_a^c$ where the second term is positive and independent of $H_E$. $H_\mathrm{c0}, H_\mathrm{SF}$ being approximately independent of pressure requires that they do not change as $H_E$ and $H_a$ are varied by $p$. Based on the above, their variations approximately relate as
\begin{equation}
\label{eq:dHE2}
    \delta H_\mathrm{c0(SF)} \propto \delta H_E + c\cdot \delta H_a 
\end{equation}
where $c>0$. Therefore, the independence of pressure, i.e. $\delta H_\mathrm{c0(SF)} =0$ requires that $\delta H_E$ and $\delta H_a$ have opposite sign. In conclusion, Eqs.\,\ref{eq:dHE1}, \ref{eq:dHE2} lead to $\delta H_E>0$ and $\delta H_a <0$ as summarized in the main text.

\subparagraph{Quantitative evaluation of $H_E, H_a$ versus pressure:} \mbox{}
\vspace{0.1cm}

The above appraisal is well matched by quantitative analysis. In Fig.\,\ref{fig:linchain_fit}(a), besides the experimental transition fields (solid markers), we also plot their fits (empty markers) based on the analytically predicted phase boundaries in Fig.\,\ref{fig:linchain}(d). The model parameters $H_E$, $H_a$ produced by the fits are plotted in Fig.\,\ref{fig:linchain_fit}(b). The former increases by approximately 25\% with pressure, while the latter decreases by about 40\%. We have also plotted the experimental fields in Fig.\,\ref{fig:linchain_fit}(c) (solid markers) over the theoretical phase diagram (solid lines), i.e. they have been rescaled by $H_\mathrm{FM0}$. Pressure reduces the ratio $H_a / H_a^c$ by half, moving all points lower on the graph. We emphasize that a larger value of $H_\mathrm{FM} (2\,\mathrm{GPa})$ produces an even more significant change in the parameters. 

Since $H_a / H_a^c$ is so low, based on the simulated colormap of Fig.\,\ref{fig:linchain}(d) we expect a negligible hysteresis $\delta$ at the spin-flop (AFM-cAFM) transition, which is consistent with the Hall measurements in the main text.

\FloatBarrier 

\end{document}